\documentclass[10pt]{iopart}
\expandafter\let\csname equation*\endcsname\relax
\expandafter\let\csname endequation*\endcsname\relax
\usepackage{graphicx}% Include figure files
\usepackage{lipsum}
\usepackage{bm}% bold math
\usepackage{tabu}
\usepackage[utf8]{inputenc}
\usepackage[T1]{fontenc}
\usepackage{mathptmx}
\usepackage{color}
\usepackage{xcolor}
\usepackage{cite}
\usepackage{iopams}  
\usepackage{amsmath}
\newcommand{\bl}[1]{\textcolor{blue}{ #1}}
\begin{document}
\title[High-speed generation of vector beams through random spatial multiplexing]{High-speed generation of vector beams through random spatial multiplexing}

\author{Xiao-Bo Hu$^1$, Si-Yuan Ma$^1$, Carmelo Rosales-Guzm\'an$^{1,2}$}
\address{$^1$ Wang Da-Heng Collaborative Innovation Center, Heilongjiang Provincial Key Laboratory of Quantum Manipulation and Control, Harbin University of Science and Technology, Harbin 150080, China.}
\address{$^2$ Centro de Investigaciones en \'Optica, A.C., Loma del Bosque 115, Colonia Lomas del campestre, C.P. 37150 Le\'on, Guanajuato, Mexico}
\ead{carmelorosalesg@cio.mx,carmelorosalesg@hrbust.edu.cn}

%%%%%%%%%%%%%%%%%%%%%%%%%%%%%%%%%%%%%%%%%%%%%%%%%%%%%%%%%%%%%%%%%%%%%%
\begin{abstract}
Complex vector modes have become topical of late due to their fascinating properties and the many applications they have found across a broad variety of research fields. Even though such modes can be generated in a wide variety of ways, digital holography stands out as one of the most flexible and versatile. Along this line, Digital Micromirror Devices (DMDs) have gain popularity in recent time due to their high refresh rates, which allows the generation of vector modes at kHz rates. Nonetheless, most techniques are limited either by the diversity of vector modes that can be generated or by the speed at which they can be switched. Here we propose a technique based on the concept of random encoding, which allows the generation of arbitrary vector beams at speeds limited only by the refresh rate of the DMD. Our technique will be of great relevance in research fields such as optical communications, laser material processing and optical manipulation, amongst others.   
\end{abstract}
 
\ioptwocol
\maketitle
%%%%%%%%%%%%%%%%%%%%%%%%%%%%%%%%%%%%%%%%%%%%%%%%%%%%%%%%%%%%%%%%%%%%%%
\section{Introduction}

Complex vector light fields, also known as vector or classically-entangled modes, have captured the attention of many researchers across a vast variety of research fields, such as optical tweezers, super resolution microscopy, classical and quantum communications, optical metrology, amongst others \cite{Rosales2018Review,Roadmap,Zhan2009,Bhebhe2018,BergJohansen2015,Hu2019}. One of the main features of vector modes is their non-homogeneous polarisation distributions, which arises from the non-separability between the spatial and polarisation degrees of freedom \cite{Galvez2012,Beckley2010,Galvez2015,Otte2016}. Importantly, their mathematical structure is identical to that of quantum entangled states, one of the main reasons why the term classically-entangled modes has been adopted to refer to these modes \cite{konrad2019quantum,forbes2019classically,toninelli2019concepts}. This mathematical similarity allows to establish many connections between the classical and the quantum world, connections that have been the subject of an extensive research that includes a wide variety of applications \cite{Balthazar2016,Silva2016,Eberly2016,Li2016,Qian2017,Borges2010,Ndagano2017,Toppel2014}. This has, in part, boosted the need for novel techniques capable to generate classically-entangled modes in easier ways, with more flexibility or with high switch speeds. Even though there are already a wide variety of techniques (see for example \cite{Rosales2018Review,Ndagano2018}), not all of them offer the flexibility and speed that is required in applications such as optical communications or laser material processing. This is the case of Spatial Light Modulators (SLMs), which offer a great flexibility for the generation of arbitrary vector modes, but are limited to low refresh rates ($\approx$ 60 Hz) \cite{Dudley2013,Chen2014,Wang2007,Maurer2007,Moreno2012,SPIEbook,Rosales2017}. In contrast, Digital Micromirror Devices (DMDs) can also generate arbitrary vector modes at higher refresh rates \cite{Ren2015,Mitchell2016,Scholes2019,Gong2014,Yao-Li2020}. In addition, DMDs are not sensitive to polarisation, a property that has been fully exploited not only in the generation but also in the characterisation of vector modes \cite{Zhaobo2020,Rosales2020,Selyem2019,Manthalkar2020,Zhaobo2019}.

While many DMD-based techniques have been developed to generate light modes with homogeneous polarisation distribution (scalar modes), only a few have been proposed to generate vector modes. In one of the first approaches, which relies on the combination of a DMD and a q-plate, a scalar mode was generated from the DMD and subsequently passed through a q-plate to convert it into a vector mode \cite{Gong2014}. Alternative techniques rely on the independent manipulation of both degrees of freedom. For this, two orthogonal spatial modes with orthogonal polarisation components are generated along independent paths and subsequently recombined along a common propagation path, using for example beam displacers \cite{Mitchell2016}. It is worth mentioning that this approach is limited to the generation of Laguerre-Gaussian ($LG_p^\ell$) modes since they can be obtained as a linear superposition of Hermite-Gaussian ($HG_{nm}$) modes\cite{Beijersbergen1993,Oneil2000}. For example, to generate a radially polarised $LG_0^1$ vector mode, a horizontally polarised $HG_{10}$ mode is generated along one path and a vertically polarised $HG_{01}$ mode on the other path. In a more recent approach, which relies on the polarisation-independence property of DMDs, the interferometric recombination was avoided by illuminating the DMD with two orthogonally-polarised modes impinging at different angles. Nonetheless, this technique is also limited to the set of $LG_p^\ell$ vector modes \cite{Rosales2020,Selyem2019}. Finally, it is worth mentioning that some DMDs can also display gray-scale holograms, at the cost of reduced refresh rates, allowing the generation of vector modes with other symmetries, this by encoding multiplexed gray-scale hologram \cite{Yao-Li2020}.

Even though, research fields such as optical tweezers, free-space optical communications, laser material processing, amongst others, will certainly benefit from vector modes with arbitrary shapes and high speeds, current techniques does not meet both requirements. As such, in this manuscript, we propose a novel technique capable to generate arbitrary vector modes with almost any spatial shape reaching speeds limited only by the refresh rate of the specific DMD (9.5 KHz for the DLP LightCrafter evaluation module). Our technique relies on the concept of random spatial multiplexing, which was proposed almost three decades ago in the context of phase-only optical filters for optical pattern recognition \cite{Davis1994} and recently in the context of beam shaping with phase-only diffractive holograms \cite{RosalesGuzman2013Airy,MartinezFuentes2018}. 

The manuscript is organised as follows, we will start by briefly explaining the main idea behind the generation of scalar beams with DMDs. Afterwards, we will focus on the concept of random spatial multiplexing and its application in the generation of two independent scalar beams, which is key to generate arbitrary vector beams at high speeds. Finally, we will demonstrate the experimental generation of vector modes with elliptical coordinates.

%%%%%%%%%%%%%%%%%%%%%%%%%%%%%%%%%%%%%%%%%%%%%%%%%%%%%%%%%
\section{Theoretical background}
\subsection{Beam shaping with digital micromirror devices}
A DMD consist of an array of millions of micron-sized mirrors ($\approx 8 \mu m$), each of which can be turned to an OFF or ON state by tilting it $-12^\circ$ or $+12^\circ$, respectively. In this way, when the DMD is properly aligned, each mirror in the ON state will reflect light in the desired direction  \cite{Scholes2019,Mirhosseini2013}. Since each mirror can only be in two states, it is important to have in mind that one of the main requirements is to address the DMD with binary holograms. Nonetheless, some devices can also deal with grays-scale holograms, at the expense of reduced refresh rates by allowing the micromirrors to oscillate between the ON and OFF states.

\begin{figure}[tb]
   \centering
    \includegraphics[width=0.49\textwidth]{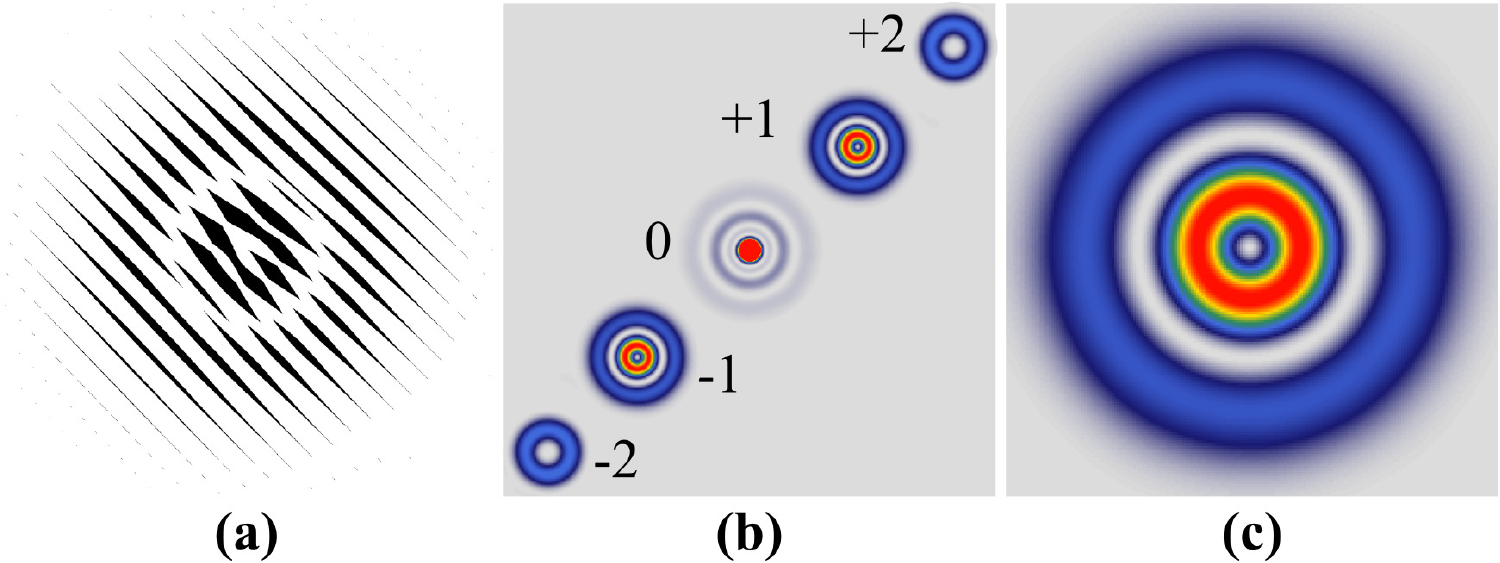}
    \caption{(a) Example of a binary hologram for the case of an $LG_1^{-1}$ mode, (b) diffraction patter produced when such hologram is illuminated by a flat wave front, here we only show the five diffraction orders. (c) Intensity profile of the positive first diffraction order, which corresponds to the encoded mode $LG_1^{-1}$.}
   \label{DifOrder}
\end{figure}

To start with, let us remind that binary-amplitude holograms are capable to shape both the amplitude and phase of a complex light field $U(x,y)=A(x,y)\exp(i\phi)$, where $\phi$ is the azimuthal angle given as $\arctan(y/x)$. To this end, a binary amplitude hologram is appropriately designed to carry in its first diffraction order the amplitude and phase information. Further, the angle of the first diffraction order, and therefore its position in the observation plane, can be controlled by adjusting the spatial frequency of the grating. The desired amplitude $A(x,y)$ at a given position $(x,y)$ in the observation plane is then acquired by locally varying the width of each diffraction grating. In a similar way, the desired phase $\exp(i\phi)$ is achieved by locally varying the lateral position of each grating. More explicitly, the binary transmittance function $T(x,y)$ is given by  \cite{Mirhosseini2013,Mitchell2016},
\begin{equation}
T(x,y)=\frac{1}{2}+\frac{1}{2}\text{sgn}\left\{\cos\left[p(x,y)\right]+\cos\left[q(x,y)\right]\right\},
\label{TM}
\end{equation}
with,
\begin{equation}
\begin{split}
p(x,y)&=\phi(x,y)+2\pi(\nu x+\eta y),\\
q(x,y)&=\arcsin\left({A(x,y)}/{A_{ max}}\right).
\end{split}
\end{equation}
Here, sgn$\{\cdot\}$ is the sign function and ${A_{max}}$ represents the maximum value of $A(x,y)$. The term $2\pi(\nu x+\eta y)$ is a linear phase ramp of spatial frequencies $\nu$ and $\eta$, that determines the angle of the first diffraction order. Figure \ref{DifOrder} (a) shows an example of a typical binary hologram obtained from Eq. \ref{TM} for the specific case of Laguerre-Gaussian ($LG_p^\ell$) modes, as defined in \cite{Hu2020}, with radial index $p=1$ and azimuthal index or topological charge $\ell=-1$. In this and all the following figures showing a binary hologram, were inverted the color in such away that the black color represents the mirrors in the ON state, which are the ones that redirect light in the desired direction. When such transmission grating is displayed on a DMD, we have to use the negative of such image. Notice how the width of each grating (the black apertures) varies across the transverse plane, resembling the amplitude of the $LG_1^{-1}$ mode. In addition, the grating clearly shows the typical fork hologram, which is commonly used to generate light beams with an azimuthal phase variation of the form $\exp(-i\ell\phi)$, in this case $\ell=-1$. Figure \ref{DifOrder} (b) schematically illustrates, through a numerical simulation, the effect of such holograms when illuminated with an expanded light beam. As expected, several diffraction orders appear, five of which are shown here. As can be seen, the first diffraction order carries the desired light field $U(x,y)=A(x,y)\exp(i\phi)$, which for the sake of clarity is shown in Fig. \ref{DifOrder} (c). In the following section we will explain the concept of random spatial multiplexing applied to the simultaneous generation of two spatial modes with a unique properties, such as, phase, amplitude or diffraction angle. 
%%%%%%%%%%%%%%%%%%%%%%%%%%%%%%%%%%%%%%%%%%%%%%%%%%%%%%%%%%%%%%%%%%%%%%%
\subsection{Random binary encoding}
\begin{figure}[tb]
   \centering
    \includegraphics[width=0.49\textwidth]{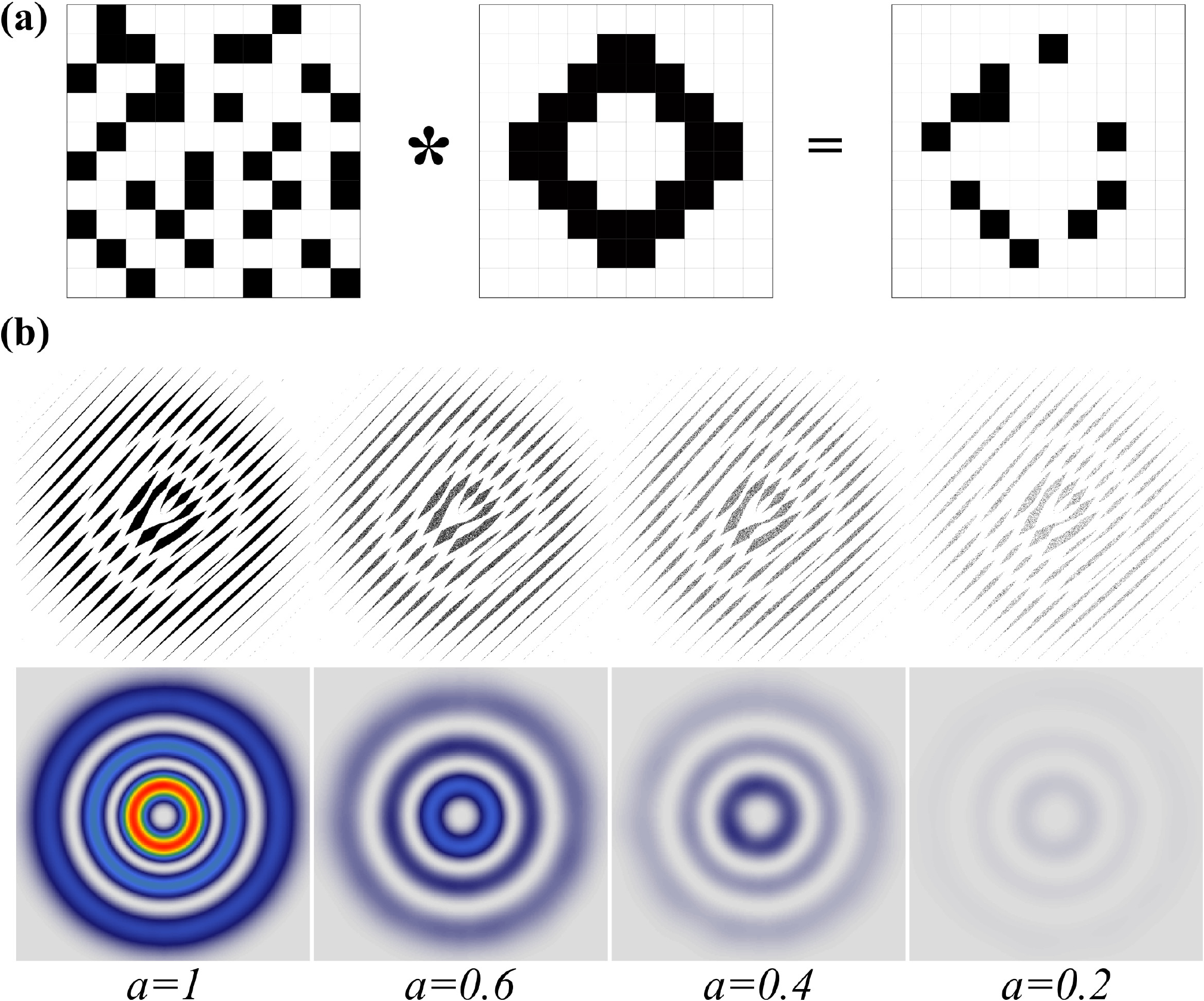}
    \caption{(a) A random binary mask (left) and a transmittance function (middle) are convoluted to obtain the the mask shown on the right. (b) Transmittance function (top) and associated intensity profile (bottom) obtained for an $LG_2^2$ for the specific values $a=1$ $a=0.6$, $a=4$ and $a=0.2$.}
   \label{RandEnc}
\end{figure}

As mentioned earlier, the concept of spatial random multiplexing has been used before in the context of light beam shaping with phase-only holograms. In particular, in \cite{RosalesGuzman2013Airy}, it was used to generate two Airy-vortex beams simultaneously. Here, we will implement a similar approach where the concept of spatial random multiplexing will be applied to multiplex two spatial modes on a single binary amplitude hologram. To start with, we define a random binary mask $R(x,y;a)$ having the same resolution as the DMD (1920$\times$1080 pixels), which takes values 0 or 1 at random pixels, as schematically shown in the left panel of Fig. \ref{RandEnc}(a), for a 10$\times$10 mask. Here, $a\in [0, 1]$ is a parameter that controls the amount of pixels in the ON state. For example, when $a=1$ all the pixels are in the ON state, whereas for $a=0.5$ only half of the them are in the ON state. In the example shown in Fig. \ref{RandEnc}(a) only one third of the pixels are in the ON state ($a=0.32$). When such random mask is displayed on the DMD, it has the effect of tilting the micromirrors to the ON or to the OFF state. Remember that only those in the ON state will diffract light in the desired direction (the pixels represented by black color). In this way, when a transmittance function $T(x,y)$ (see middle panel of Fig. \ref{RandEnc} (a)) is convoluted with the random binary mask $R(x,y;a)$, some of its pixels in the ON state are randomly switched to the OFF state (see right panel of Fig. \ref{RandEnc} (a)). To further exemplify this, let us take the specific case of $LG_p^\ell$ modes, which is illustrated in Fig. \ref{RandEnc} (b). The top panels show the effect of multiplying the transmittance function associated to an $LG_2^2$ mode by a random binary mask $R(x,y;a)$, as function of $a$, here we show the cases $a$=1, 0.8, 0.6 and 0.2. Notice that a decrease of the parameter $a$ is associated to a decrease in the number of micromirrors that contribute to the formation of the beam in the first diffraction order. This decrease results in a decrease of its power, as illustrated in the bottom panels of Fig. \ref{RandEnc} (b). Hence the parameter $a$ provides a way to directly control the power of the mode associated to the transmittance function $T(x,y)$. In the following section, we will further explore this idea in the simultaneous generation of two light beams with different powers.

\subsection{Spatial random multiplexing}
\begin{figure}[tb]
    \centering
    \includegraphics[width=0.49\textwidth]{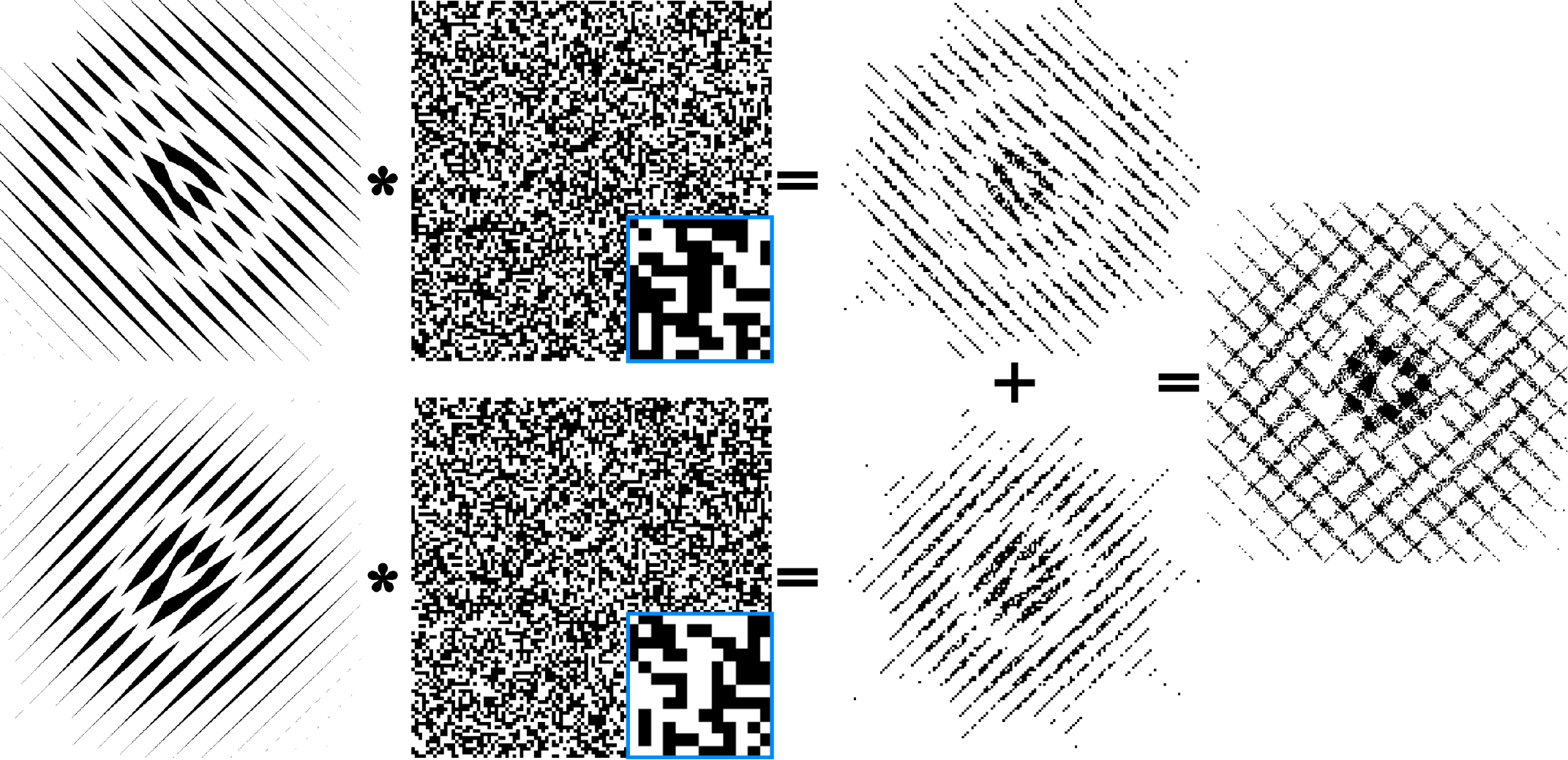}
    \caption{Conceptual representation of spatial random encoding. The left column illustrates an example for the transmittance function $T_1(x,y)$ (top) and $T_2(x,y)$ (bottom), encoding the modes $LG_2^{-1}$ and $LG_1^{1}$. The second column show an example of a random binary mask $R(x,y;a)$ (top) and its complementary random binary mask $\Bar{R}(x,y;1-a)$ (bottom), the small inset show an enlarged portion of the mask to emphasize that one is the complement of the other. The third column shows the convolution $T_1(x,y)*R(x,y;a)$ (top) and $T_2(x,y)*\Bar{R}(x,y;1-a)$ (bottom). The last panel shows the resultant transmittance function $T_f(x,y)$.}
    \label{concept}
\end{figure}

We will now explain how the approach described above allows the simultaneous generation of two spatially independent modes, each with unique amplitude, phase, diffraction angle and power content. We will start by defining the complementary function $\Bar{R}=1-R(x,y;1-a)$ with the property $R(x,y;a)\Bar{R}(x,y;1-a)=0$. This implies that for $a=1$, all the pixels of $R(x,y;1)$ are in the On state, whereas for $\Bar{R}(x,y;0)$ all are in the OFF state. On the contrary, for $a=0$, all the pixels of $R(x,y;0)$ are in the OFF state, while for $\Bar{R}(x,y;1)$ all are in the ON state. The next step is to perform the convolution of $R(x,y;a)$ with $T_1(x,y)$ and $\Bar{R}(x,y;1-a)$ with $T_2(x,y)$ and multiplex this two holograms into a single one. This mathematically expressed as, 
\begin{equation}
T_{f}(x,y)=R(x,y;a)*T_1(x,y)+\Bar{R}(x,y;1-a)*T_2(x,y).
\label{TMultiplexed}
\end{equation}
\noindent
This transmittance function will generate two modes simultaneously with some of the pixels contributing to the generation of the mode associated to $T_1(x,y)$ and the rest to the mode encoded in $T_2(x,y)$. It is important to mention that even when some of the pixels of $T_1(x,y)$ overlap with some of the pixels of $T_2(x,y)$, this overlaps disappears after the convolution with the binary masks $R(x,y;a)$ and $\Bar{R}(x,y;1-a)$. The previous description is schematically illustrated in Fig.\ref{concept} for the specific case of $LG_p^\ell$. The first column shows an example of the transmittance function $T_1(x,y)$ (top) and $T_2(x,y)$ (bottom). The second column show an example of a random binary mask $R(x,y;a)$ (top) and its complementary random mask $\Bar{R}(x,y;1-a)$ (bottom). The inset shows an enlarged portion of both masks to emphasize one is the complement of the other. The third row shows the convolution $T_1(x,y)*R(x,y;a)$ (top) and $T_2(x,y)*\Bar{R}(x,y;1-a)$ bottom, which are finally added to produce the transmittance function $T_f(x,y)$ shown on the right most inset of the same figure. For this specific example, we encoded the $LG_2^{-1}$ mode in the function $T_1(x,y)$ and $LG_1^1$ mode in the function $T_2(x,y)$. 

\begin{figure}[tb]
   \centering
    \includegraphics[width=0.49\textwidth]{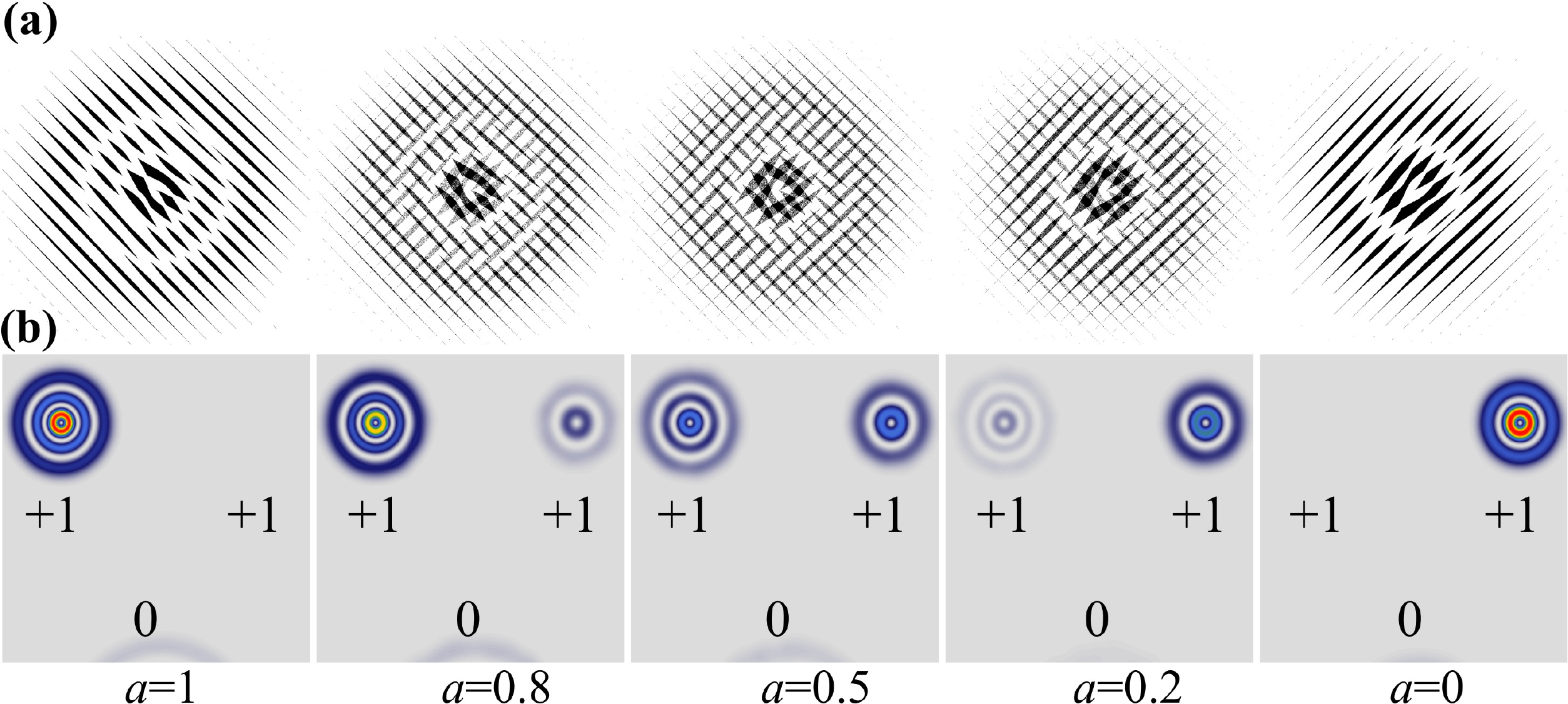}
    \caption{(a) Transmittance function $T_f(x,y)$ and (b) Far field intensity distribution of as function of the parameter $a$ for $a=1$, 0.8, 0.5, 0.2 and 0. The examples shown correspond to the modes $LG_2^{-1}$ (left) and $LG_1^{1}$ (right).}
   \label{TwoBeams}
\end{figure}

We further illustrate this technique by using as an example the set of $LG_p^\ell$ modes. More specifically, by generating simultaneously the modes $LG_2^{-1}$ and $LG_1^{1}$. First, in Fig. \ref{TwoBeams} (a) we show the transmittance function $T_f(x,y)$ as function of the parameter $a$ for $a=1, 0.8, 0.5, 0.2$ and $0$. In a similar way, in Fig.  \ref{TwoBeams} (b) we show the intensity distribution of the two generated modes. Notice that for $a=1$ only the hologram corresponding to the $LG_2^{-1}$ mode appears, as $a$ decreases, the hologram corresponding to the mode $LG_1^{1}$ starts appearing, while the first one gradually vanishes until it completely disappears for $a=0$. In Fig. \ref{TwoBeams} (b) we show the far-field intensity profile produced by the holograms shown in Fig. \ref{TwoBeams} (a), where the first diffraction order is shown and the 0$^{th}$ is only indicated for reference purposes. As expected, for $a=1$ only the $LG_2^{-1}$ mode is present, as we decrease $a$, its power decreases while the power of the $LG_1^1$ mode increases, for $a=0.5$ the power of both beams is the same. Finally for $a=0$, only the mode $LG_1^1$ is present.
%%%%%%%%%%%%%%%%%%%%%%%%%%%%%%%%%%%%
\subsection{Generation of vector modes through spatial random multiplexing}
To start this section, let us remind that vector modes result from a non-separable coupling between the spatial and polarisation degrees of freedom. Experimentally, such modes can be generated as a coaxial superposition of two orthogonal spatial modes carrying orthogonal polarisation. Mathematically, this can be expressed as \cite{Galvez2012,Galvez2015}
\begin{equation}\label{Eq:VM}
    {\bf U}({\bf r})=\cos(\theta) U_1({\bf r})\hat{\bf e}_1+\sin(\theta) \e^{i\alpha}U_2({\bf r})\hat{\bf e}_2,
\end{equation}
where $U_1({\bf r})$ and $U_2({\bf r})$ are two normalized complex-valued orthogonal functions representing the spatial degree of freedom, whereas $\hat{\bf e}_1$ and $\hat{\bf e}_2$ are two orthogonal vectors representing the state of polarisation. The coefficient $\theta\in[0,\pi/2]$ allows a transition between scalar ($\theta=0$ and $\theta=\pi/2$) and vector ($\theta=\pi/4$). The complex term $\exp(i\alpha)$, with $\alpha\in [0, \pi]$, represents an inter-modal phase between both constituting spatial modes. 

We will explain how to use random binary encoding to generate vector beams with arbitrary spatial shape and polarisation distribution. The main idea behind this technique relies on the ability to generate two independent and orthogonal spatial modes whose power content can be controlled also independently. The requirement of having two independent modes is essential as we also need to manipulate their polarisation state independently, as required by Eq. \ref{Eq:VM}. Once generated, these modes are recombined along a common propagation axis where the desired vector beam is generated. To achieve this superposition we will use a recently proposed method, which consist on illuminating the DMD with two modes of orthogonal polarisation impinging at different angles but exactly on the same spot, which coincides with the geometric center of the binary hologram. On the DMD we display a multiplexed hologram encoding two orthogonal spatial modes. Here is where our random spatial multiplexing technique becomes crucial since it allows us to generate two independent modes with controllable power using the transmittance function $T_f$ given by Eq. \ref{TMultiplexed}. In this way, when the two beams with orthogonal polarisation impinge on the DMD, two copies of the modes encoded in $T_f$ are generated in the first diffraction order, one with one polarisation, lets say right circular, and the other with the orthogonal polarisation, in this case left circular. All we have to do now is to adjust the grating periods of each mode to overlap two of the four modes along the same propagation axis. More details of this procedure will be provided in the following section.

%%%%%%%%%%%%%%%%%%%%%%%%%%%%%%%%%%%%%%%%%%%%%%%%%%%%
\section{Experimental generation of vector modes through spatial random multiplexing}
\subsection{Experimental setup}
\begin{figure*}[tb]
    \centering
    \includegraphics[width=0.97\textwidth]{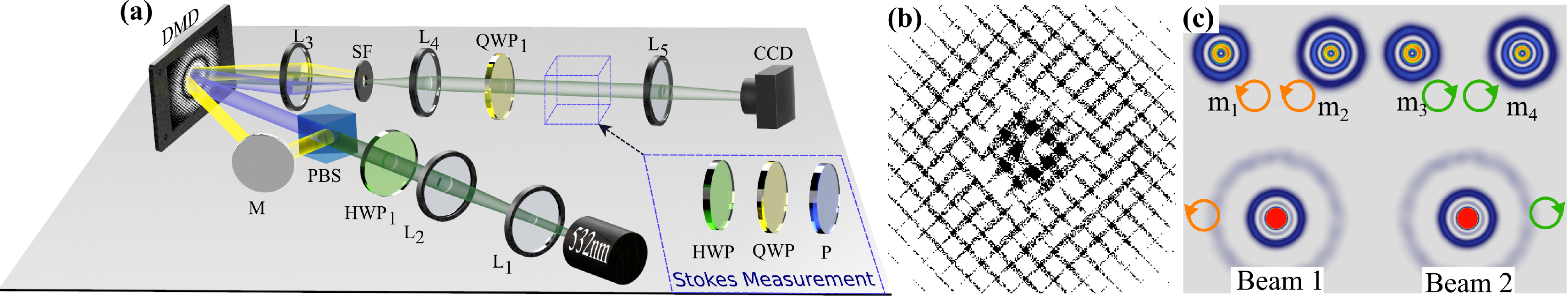}
    \caption{(a) Experimental setup implemented to generate vector modes using binary random encoding. A horizontally polarized laser beam, expanded and collimated with lenses L$_1$ and L$_2$ is changed to diagonal polarisation with a Half-Wave Plate (HWP$_1$) oriented at 45$^\circ$ and separated afterwards into its two polarisation components using a Polarising Beam Splitter (PBS). Both beams are then sent to the centre of a Digital Micromirror Device (DMD), where the hologram is displayed. A Spatial Filter (SF) placed in the focal plane of a telescope formed by lenses L$_3$ and L$_4$ filters the desired order. A Quarter-Wave Plate (QWP$_1$) at 45$^\circ$ ensures the vector mode is in the circular polarisation basis, which is imaged with lens L$_5$ to a Charge-Coupled Device (CCD) camera. The state of polarisation is reconstructed through Stokes polarimetry by means of a HWP, a QWP and a polariser (P). (b) Typical hologram displayed on the DMD. (c) Intensity profile of the first diffraction order generated by Beam1 and Beam2, which generates modes m$_1$, m$_2$ and m$_3$, m$_4$, respectively, with orthogonal polarisation, represented by the orange and green circular arrows.}
    \label{setup}
\end{figure*}

To generate the modes described by Eq. \ref{Eq:VM} we will use the approach described in \cite{Rosales2020}, based on the experimental setup depicted in Fig. \ref{setup}(a). Here, a horizontally polarised laser beam ($\lambda=532$ nm, 300 mW) is first expanded with the pair of lenses L$_1$ and L$_2$ of focal lengths $f_1=20$ mm and $f_2=200$ mm, respectively. The beam is then sent through a Half-wave plate (HWP) oriented at $22.5^\circ$, to obtain a diagonally polarised beam. Afterwards it is split into its horizontal and vertical polarisation components using a Polarising Beam Splitter (PBS). Both beams are then directed to a polarisation-independent Digital Micromirror Device (DMD, DLP Light Crafter 6500 from Texas Instruments), impinging at slightly different angles ($\approx 1.5^\circ$) at the center of the DMD. Here, a spatial random multiplexed hologram is displayed, which is generated as explained before using the transmittance function given by Eq. \ref{TMultiplexed}. The hologram contains the two constituting wave fields $U_1(x,y)$ and $U_2(x,y)$ required to generate the vector mode given by Eq.\ref{Eq:VM}, which is shown in Fig. \ref{setup}(b). As result, four modes will appear in the first diffraction order, two for each beam impinging on the DMD, as shown in Fig. \ref{setup}(c), where modes m$_1$ and m$_2$ correspond to Beam 1 and modes m$_3$ and m$_4$ to Beam 2. Crucially, modes  m$_2$ and m$_3$, which are already orthogonal, also carry orthogonal polarisation, as indicated in Fig. \ref{setup}(c). Hence, all we have to do is to adjust the period of the diffraction grating of each beam to overlap them along a common propagation axis. The resulting mode is already the desired vector beam, which we isolate with the help of a Spatial Filter (SF), located at the focusing plane of a telescope composed by lenses $\rm L_3$ and $\rm L_4$, both with focal lengths $f=100$ mm. A Quarter-Wave plate (QWP$_1$) can be added either before or after the DMD to change the vector mode from the linear to the circular polarisation basis. Finally, we add a lens $\rm L_5$ of focal length $f=300$ mm to send the generated modes to a Charge-Coupled Device (CCD) camera (FLIR FL3-U3-120S3C-C from Pointgray).

\subsection{Polarisation reconstruction}
One way to analyse the generated vector modes is by reconstructing their state of polarisation through the Stokes parameters, which can be computed from a series of intensity measurements as \cite{Goldstein2011}, 
\begin{equation}\label{Eq.Stokes}
\begin{split}
\centering
 &S_{0}=I_{0},\hspace{19mm} S_{1}=2I_{H}-S_{0},\hspace{1mm}\\
 &S_{2}=2I_{D}-S_{0},\hspace{10mm} S_{3}=2I_{R}-S_{0},
\end{split}
\end{equation}
Here $I_0$ is the total intensity of the beam, $I_H$, $I_D$ and $I_R$ are the corresponding intensities of the horizontal, diagonal and right-handed polarisation components. Experimentally, these intensity are measured using a series of polarisation filters, composed by a linear polariser (P), a HWP and a QWP, as indicated in Fig. \ref{setup}, and recorded with the CCD camera  \cite{Zhaobo2019,Manthalkar2020}. More precisely, the intensities of the horizontal ($I_H$) and diagonal ($I_D$) polarisation components, are obtained by inserting a linear polariser at $0^\circ$ and $45^\circ$, respectively, whereas the intensity corresponding to the right circular polarisation component ($I_R$) is acquired by inserting a QWP at $45^\circ$ in combination with a linear polariser at $90^\circ$.

\subsection{Experimental generation of Laguerre-Gaussian vector beams}
\begin{figure}[tb]
   \centering
    \includegraphics[width=0.49\textwidth]{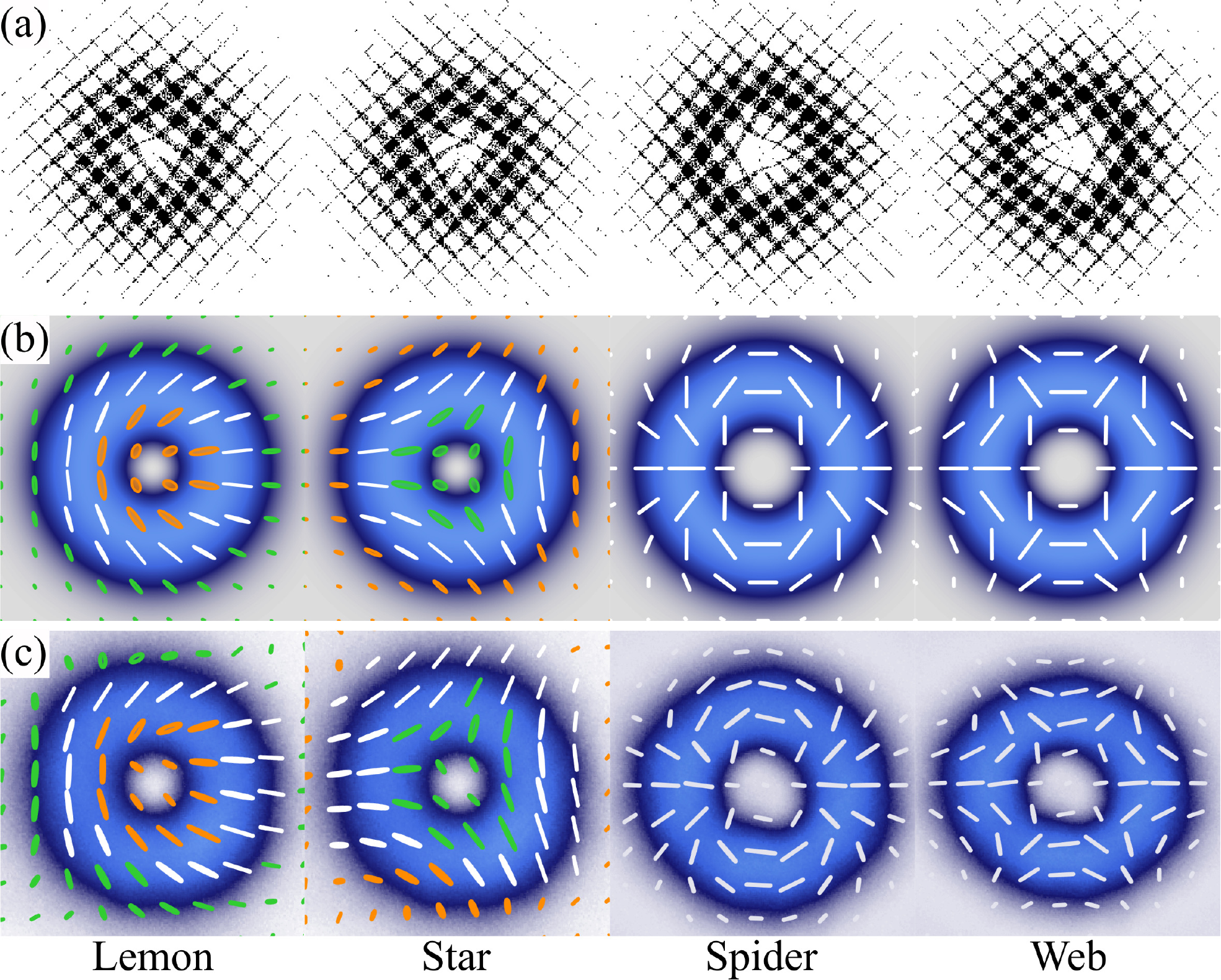}
    \caption{Experimental generation of the Laguerre-Gaussian vector modes $\{LG_0^1;LG_0^2\}$ (Lemon), $\{LG_0^2;LG_0^1\}$ (Start), $\{LG_0^{-2};LG_0^2\}$ (Spider) and $\{LG_0^{2};LG_0^{-2}\}$ (Web). (a) Examples of the holograms displayed on the DMD, where, for the purpose of display, we inverted the colors. (b) Numerical simulation of transverse intensity profile overlapped with its corresponding polarisation distribution. (c) Experimentally generated modes.}
   \label{LGBeams}
\end{figure}

As a first example we will show the arbitrary generation of cylindrical vector modes encoded in the $LG_0^\ell$ spatial modes. For the sake of clarity, we rewrite Eq. \ref{Eq:VM} using this spatial basis and the circular polarisation basis, $\hat{\bf e}_R$ and $\hat{\bf e}_L$,
\begin{equation}
{\bf U}_{LG}({\bf r})=\cos\theta LG_{p_1}^{\ell_1}({\bf r})\hat{\bf e}_R+\sin\theta \e^{i\alpha}LG_{p_1}^{\ell_2}({\bf r})\hat{\bf e}_L.
\label{LGvector}
\end{equation}
Examples of the typical vector beams  that can be generated are shown in Fig. \ref{LGBeams} for the specific cases $\{LG_0^1;LG_0^2\}$ (Lemon), $\{LG_0^2;LG_0^1\}$ (Star), $\{LG_0^{-2};LG_0^2\}$ (Spider) and $\{LG_0^{2};LG_0^{-2}\}$ (Web). An example of the multiplexed holograms required to generate these modes is shown in Fig. \ref{LGBeams} (a). In a similar way, Fig. \ref{LGBeams} (b) shows a theoretical simulation of the transverse intensity profile, overlapped with the transverse polarisation distribution. Finally, Fig. \ref{LGBeams} (c) shows the experimental intensity profile, also overlapped with the reconstructed polarisation distribution. Small deviation between the numerical simulations and our experimental results are to be expected since all DMDs introduce optical aberrations caused by the non-flatness of the screen, nonetheless, they can be minimized by adding a correction mask \cite{Scholes2019}. 

\begin{figure}[tb]
   \centering
    \includegraphics[width=0.49\textwidth]{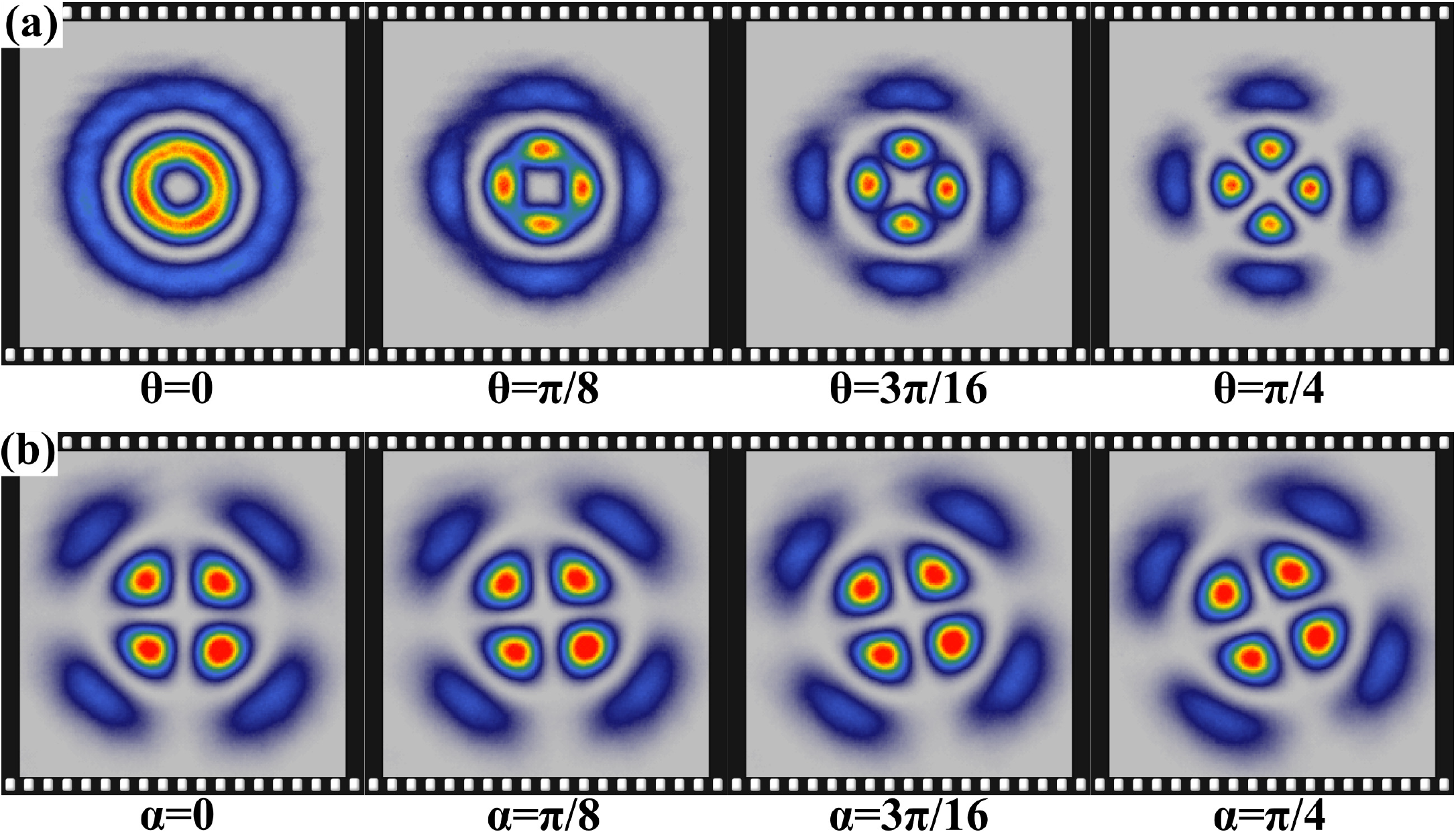}
    \caption{(a) Selected frames taken from \bl{Media 1} where the intensity distribution of the vector modes as function of $\theta\in[0,\pi/2]$ is shown. (b) Selected frames taken from \bl{Media 2} where the intensity distribution as function of $\alpha\in[0,\pi]$ is shown. In both cases, the vector beam was sent through a linear polarise oriented horizontally.}
   \label{LG2}
\end{figure}

As a second example, we show the case of Laguerre-Gaussian vector beams with arbitrary values of $\theta$ and $\alpha$. Without loss of generality, we restrict this example to the vector mode formed by the superposition of the modes $LG_1^2$ and $LG_1^{-2}$. As mentioned earlier, $\theta$ modulates the contribution of each spatial mode, which we achieved experimentally by varying the coefficient $a$ in the transmittance function associated to each mode, namely, $R(x,y;a)*LG_1^2$ and $\Bar{R}(x,y;1-a)*LG_1^{-2}$. The intensity profile of modes generated for increasing values of $a\in[0,1]$ ($\theta\in[0,\pi/2$]), are shown as a media file (\bl{Media 1}), selected frames this video are shown in Fig. \ref{LG2} (a). In order to see the effect produced when $\theta$ is increased, we passed the vector beam through a linear polariser oriented horizontally. As result of this, the intensity profile of the mode observed with the CCD, gradually changes from two concentric rings of light (scalar), to a petal-like structure (vector) and back to the structure of concentric rings (scalar). Finally, the effect of changing the inter-modal phase associated to $\alpha$ can be achieved experimentally, by simply adding an extra phase to one of the constituting modes and encoded directly on its transmittance function. Again, the effect of $\alpha$ can be clearly observed by sending the vector beam through a linear polariser. As result, the intensity distribution features a concentric structure of four petals that will rotate anticlockwise with increasing values of $\alpha$. Figure \ref{LG2} shows a selected set of four frames extracted from \bl{Media 2}, where this effect is clearly shown. 

\subsection{Experimental generation of elliptical Ince-Gaussian beams}

Finally, and to show the capabilities of our technique, in this section we demonstrate the generation of vector modes with elliptical symmetry by encoding the spatial degree of freedom in the Helical Ince-Gaussian (HIG) vector modes. Such vector modes have been previously generated with phase holograms encoded in SLMs \cite{Otte2018a} and DMDs \cite{Yao-Li2020} and therefore with low refresh rates. Importantly, our technique allows us to generate such modes using binary holograms at speeds only limited by the specific characteristic of the DMD, in our case 9.5 kHz. Let us first remind that IG modes are natural solutions of the paraxial wave equation in elliptical cylindrical coordinates ${\bf r}=(\xi,\eta,z)$, with, $\xi\in[0,\infty)$ and $\eta\in[0,2\pi)$ representing the radial and angular elliptical coordinates, respectively \cite{Bandres2004}. Here, we will not provide specific details of how this solutions are derived, but only the mathematical expressions that describe them, more details can be found in \cite{Bandres2004}. The scalar IG mode can be classified into even or odd in pure relation to the even and odd Ince polynomials, $C_p^m(\cdot,\varepsilon)$ and $S_p^m(\cdot,\varepsilon)$, which constitute these modes. Mathematically, they have the specific form,
\begin{equation}\label{IGscalar}
\begin{split}
    IG_{q,m,\varepsilon}^e({\bf r}) &= \frac{C\omega_0}{\omega(z)}C_q^m(i\xi,\varepsilon)C_q^m(\eta,\varepsilon)\text{e}^{-\frac{r^2}{\omega(z)}}\text{e}^{-i\left(kz+Z-\Phi\right)},\\
    IG_{q,m,\varepsilon}^o({\bf r}) &= \frac{S\omega_0}{\omega(z)}S_q^m(i\xi,\varepsilon)S_q^m(\eta,\varepsilon)\text{e}^{-\frac{r^2}{\omega(z)}}\text{e}^{-i\left(kz+Z-\Phi\right)},
\end{split}
\end{equation}
where $C$ and $S$ are normalisation constants and the superscripts $e$ and $o$ stand for even and odd, respectively. Further, the subscripts $p,m\in\mathbb{N}$, which are restricted by $0\leq m\leq q$ for even functions and $1\leq m\leq q$ for odd functions, provide the order of the mode. Additional parameters, are related to the Gaussian envelope of beam waist $\omega_0$ that confines these modes. More precisely $\Phi=(p+1)\arctan(z/z_R)$ is the Gouy phase, $Z(z)=kr^2/2R(z)$ is an additional phase term related to the radius of curvature $R(z)=z+z_R^2/z$ of the phase front and $z_R=\pi\omega_0^2/\lambda$ is the Rayleigh length. In addition, the coherent superposition of the even and odd modes gives rise to the HIG modes $IG_{q,m,\varepsilon}^{h+}({\bf r})$ and $IG_{q,m,\varepsilon}^{h-}({\bf r})$, which can be expressed mathematically as,
\begin{equation}\label{helicalIG}
\begin{split}
    IG_{q,m,\varepsilon}^{h+}({\bf r}) &= IG_{q,m,\varepsilon}^e({\bf r}) + i IG_{q,m,\varepsilon}^o({\bf r}),\\
    IG_{q,m,\varepsilon}^{h-}({\bf r}) &= IG_{q,m,\varepsilon}^e({\bf r}) - i IG_{q,m,\varepsilon}^o({\bf r}),
\end{split}
\end{equation}
Perhaps, one of the most salient properties of IG modes is that they represent a larger family of modes that includes the Laguerre- and Hermite-Gaussian modes as specific cases, which is related to the ellipticity parameter, $\varepsilon=2f_0/\omega_0^2$, $\varepsilon\in[0,\infty)$. More precisely, the $IG_{q,m,\varepsilon}^{h}$ transform into the Laguerre-Gaussian ($LG_p^\ell$) when $\varepsilon=0$, with their indices related as $\ell=m$ and $p=(q-m)/2$. In a similar way, the $IG_{p,m,\varepsilon}^{e,o}$ modes become the Hermite Gaussian ($HG_{n_x n_y}$) modes for $\varepsilon\to\infty$, with their indices related as $n_x=m$, $n_y=p-m$, for even modes and $n_x=m-1$, $n_y=p-m+1$ for odd modes \cite{Bandres2004b}. Notably, the set of $IG_{q,m,\varepsilon}^{h+}({\bf r})$ modes are orthogonal to the set of modes $IG_{q,m,\varepsilon}^{h-}({\bf r})$. Hence, we can use these as the spatial degree of freedom to generate HIG vector modes. With this restriction, Eq. \ref{Eq:VM} becomes,
\begin{equation}
{\bf U}_{IG^h}({\bf r})=\cos\theta IG_{q,m,\varepsilon}^{h+}({\bf r})\hat{\bf e}_R+\sin\theta \e^{i\alpha} IG_{q,m,\varepsilon}^{h-}({\bf r})\hat{\bf e}_L
\label{IGvector}
\end{equation}

\begin{figure}[tb]
    \centering
    \includegraphics[width=0.49\textwidth]{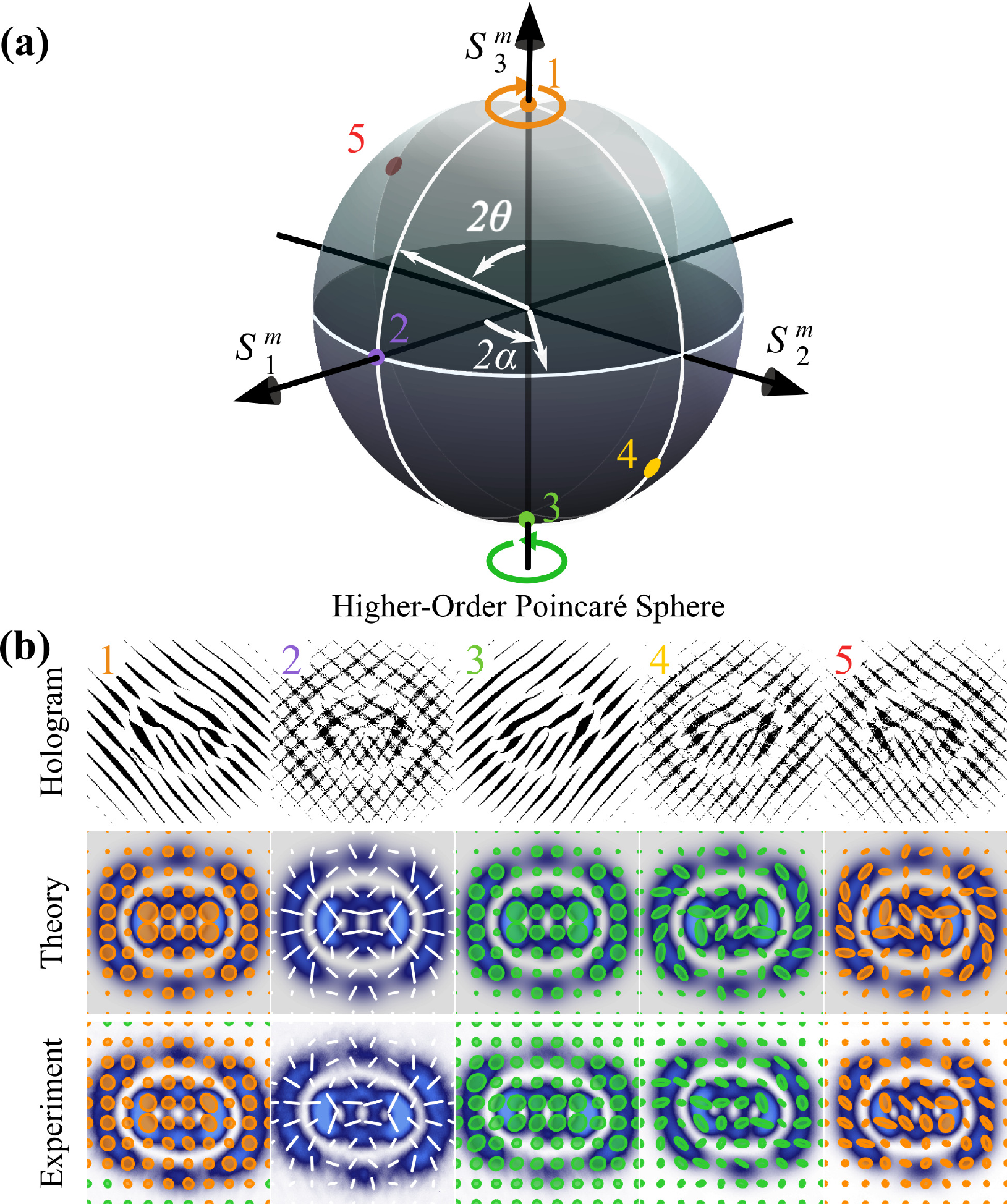}
    \caption{(a) Geometric representation of elliptical vector modes on a high order Poincar\'e sphere. (b) Typical hologram displayed on the DMD (top), numerical simulation of the intensity and polarisation distribution (middle), experimental results (bottom).}
    \label{Poincare}
\end{figure}

To begin with, we will show that our technique allows the generation of arbitrary IG vector modes as represented on a High Order Poincar\'e Sphere (HOPS) \cite{Yao-Li2020}. This geometric representation, provides a way to visualize arbitrary vector modes on the surface of a unitary sphere by associating their inter-modal phase ($\alpha$) and weighting coefficient ($\theta$) to a point with coordinates $(2\theta,2\alpha)$. In this way, the two scalar modes with right and left circular polarisation are mapped to the North and South Poles, respectively, and pure vector modes to the equator. Without loss of generality, we restrict our analysis to the specific case given by HIG vector modes ${\rm HIG}_{5,3,2}^h({\bf r})$. As a first example, Fig. \ref{Poincare} (a) shows a HOPS representation of five vector modes labeled from 1 to 5, with specific coordinates, in the same order, $(0,0), (\pi/2,0), (\pi,0), (\pi/2,3\pi/4)$ and $(3\pi/2,\pi/4)$. In this representation, the Cartesian coordinates of the HOPS are related to the higher-order Stokes parameters $S_1^{m}$, $S_2^{m}$ and $S_3^{m}$, as detailed in \cite{Milione2011}. The holograms required to generate such modes, which are also computed in the same way as explained before but with the HIG modes, are shown in the top row of Fig. \ref{Poincare} (a). Note that this holograms do not correspond to the holograms displayed in the DMD, they have been scaled for display purposes. The middle row of Fig. \ref{Poincare} (b) shows a numerical simulation of the transverse intensity profile overlapped with the polarization distribution of this specific cases. Our experimental results are shown in the bottom row of Fig. \ref{Poincare} (b), showing a high agreement with our numerical simulations.

\begin{figure}[tb]
    \centering
    \includegraphics[width=0.49\textwidth]{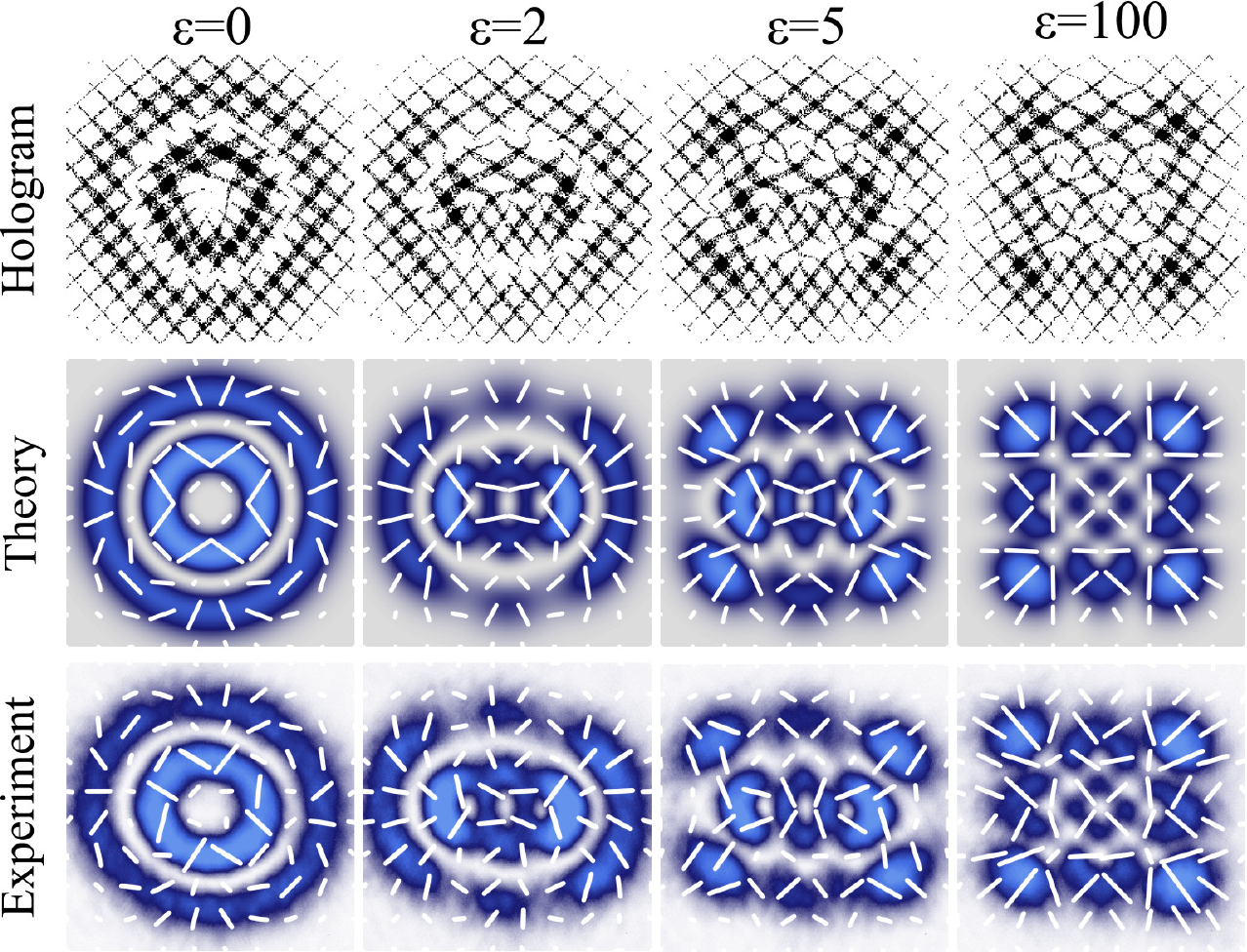}
    \caption{Helical Ince-Gaussian vector modes as a function of the ellipticity $\varepsilon$. Binary hologram(top row) numerical simulations (middle row) and experimental results (bottom row) of the transverse intensity profiles overlapped with the polarization distribution. Here, we show the specific case ${\rm IG}_{5,3,\varepsilon}^{h}$ for $\varepsilon =0, 2, 5$ and 100. Notice the transition from Laguerre- to Hermite-Gaussian vector modes.}
    \label{ec}
\end{figure}

Our technique is also capable to generate HIG vector modes with arbitrary values of ellipticity. As a way of example, Fig. \ref{ec} shows a representative set of examples for the specific case ${\rm HIG}_{5,3,\varepsilon}$ and $\varepsilon \in[0,2,5,100]$. The top row shows an example of the required holograms to generate such modes, the middle row a numerical simulation of the intensity and polarisation distribution and the third row our experimental results. Notice the high similitude between both sets.

\section{Conclusions}

Vector beams represent a general state of light in which the spatial and polarisation degrees of freedom are coupled in a non-separable way. The unique properties of these modes have become of great relevance in applications such as optical communications, laser material processing, optical metrology, amongst others. Nonetheless, most applications are based on vector modes with cylindrical symmetry, such as Laguerre- or Bessel-Gaussian modes. Noteworthy, the availability of such modes with other symmetries, such as elliptic or parabolic, will benefit current applications and pave the path towards the development of new. In addition, the generation of vector modes with high refresh rates provide an additional advantage that will certainly enhance the capabilities of these. Hence, it is almost natural to seek for novel technique which can offer high flexibility and high generation rates. Nonetheless, existing techniques capable to generate vector beams with arbitrary spatial shape are limited to low refresh rates and those with high generation rates are limited to the reduced set of cylindrical vector modes. Here, we proposed a technique that offers both advantages simultaneously. This technique is based on the concept of random binary spatial multiplexing, which in combination with the Digital Micromirror Device (DMD) technology, is capable to generate vector modes with arbitrary spatial distribution, at the maximum generation speed offered by the specific DMD model. As we demonstrated, random spatial multiplexing allow us to generate two scalar modes with independent properties, such as spatial shape, phase profile, polarisation, power content and diffraction angle. The main idea consist on using some pixels of the DMD, selected at random, to generate one scalar mode and the remaining ones to generate the other mode. These two modes are then recombined along a common propagation axis, using a polarisation-insensitive approach \cite{Rosales2020}, where the vector mode is generated. Our approach provides with a powerful tool for the generation of vector modes with arbitrary spatial shapes and high refresh rates, which will be of great relevance in various applications. For example, in the field of optical tweezers, for the localized tailoring of optical forces \cite{Bhebhe2018}, or in free-space optical communications for the generation of vector modes with high resilience to turbulence \cite{Ndagano2018} and last but not least, in laser material processing to enhance the drilling capabilities of the laser beams employed \cite{Meier2007}.

\section*{Acknowledgement} The authors would like to thank Dr. Valeria Rodriguez-Fajardo for fruitful discussions.

\section*{Funding}
This work was partially supported by the National Natural Science Foundation of China (NSFC) under Grant No.  61975047.

\section*{Disclosures}
The authors declare that there are no conflicts of interest related to this article

\section*{References}
%%%%%%%%%% If using BibTeX:
\bibliographystyle{iopart-num}
%\bibliography{References}

\begin{thebibliography}{10}
\expandafter\ifx\csname url\endcsname\relax
  \def\url#1{{\tt #1}}\fi
\expandafter\ifx\csname urlprefix\endcsname\relax\def\urlprefix{URL }\fi
\providecommand{\eprint}[2][]{\url{#2}}

\bibitem{Rosales2018Review}
Rosales-Guzm\'{a}n C, Ndagano B and Forbes A 2018 {\em J. Opt.\/} {\bf 20}
  123001

\bibitem{Roadmap}
Rubinsztein-Dunlop H, Forbes A, Berry M~V, Dennis M~R, Andrews D~L, Mansuripur
  M, Denz C, Alpmann C, Banzer P and Bauer T 2017 {\em J. Opt.\/} {\bf 19}
  013001

\bibitem{Zhan2009}
Zhan Q 2009 {\em Adv. Opt. Photonics\/} {\bf 1} 1--57

\bibitem{Bhebhe2018}
Bhebhe N, Williams P~A~C, Rosales-Guzm{\'a}n C, Rodriguez-Fajardo V and Forbes
  A 2018 {\em Sci. Rep.\/} {\bf 8} 17387

\bibitem{BergJohansen2015}
Berg-Johansen S, T\"{o}ppel F, Stiller B, Banzer P, Ornigotti M, Giacobino E,
  Leuchs G, Aiello A and Marquardt C 2015 {\em Optica\/} {\bf 2} 864--868

\bibitem{Hu2019}
Hu X~B, Zhao B, Zhu Z~H, Gao W and Rosales-Guzm\'{a}n C 2019 {\em Opt. Lett.\/}
  {\bf 44} 3070--3073

\bibitem{Galvez2012}
Galvez E~J, Khadka S, Schubert W~H and Nomoto S 2012 {\em Appl. Opt.\/} {\bf
  51} 2925--2934

\bibitem{Beckley2010}
Beckley A~M, Brown T~G and Alonso M~A 2010 {\em Opt. Express\/} {\bf 18}
  10777--10785

\bibitem{Galvez2015}
Galvez E~J 2015 {\em Light Beams with Spatially Variable Polarization\/}
  (Wiley-Blackwell) chap~3, pp 61--76 ISBN 9781119009719

\bibitem{Otte2016}
Otte E, Alpmann C and Denz C 2016 {\em J. Opt.\/} {\bf 18} 074012

\bibitem{konrad2019quantum}
Konrad T and Forbes A 2019 {\em Contemporary Physics\/}  1--22

\bibitem{forbes2019classically}
Forbes A, Aiello A and Ndagano B 2019 Classically entangled light {\em Progress
  in Optics\/} (Elsevier Ltd.) pp 99--153

\bibitem{toninelli2019concepts}
Toninelli E, Ndagano B, Vall{\'e}s A, Sephton B, Nape I, Ambrosio A, Capasso F,
  Padgett M~J and Forbes A 2019 {\em Advances in Optics and Photonics\/} {\bf
  11} 67--134

\bibitem{Balthazar2016}
Balthazar W~F, Souza C~E~R, Caetano D~P, {a}o E~F~G, Huguenin J~A~O and Khoury
  A~Z 2016 {\em Opt. Lett.\/} {\bf 41} 5797--5800

\bibitem{Silva2016}
Diego G~S, Robert B, Felix Z, Christian V, Markus G, Matthias H, Stefan N,
  Michael D, Andrea A, Marco O and Alexander S 2016 {\em Laser Photonics
  Rev.\/} {\bf 10} 317--321

\bibitem{Eberly2016}
Eberly J~H, Qian X~F, Qasimi A~A, Ali H, Alonso M~A, Guti{\'e}rrez-Cuevas R,
  Little B~J, Howell J~C, Malhotra T and Vamivakas A~N 2016 {\em Phys. Scr\/}
  {\bf 91} 063003
  \urlprefix\url{http://stacks.iop.org/1402-4896/91/i=6/a=063003}

\bibitem{Li2016}
Li P, Wang B and Zhang X 2016 {\em Opt. Express\/} {\bf 24} 15143

\bibitem{Qian2017}
Qian X~F, Vamivakas A~N and Eberly J~H 2017 {\em Opt. Photon. News\/} {\bf 28}
  34--41

\bibitem{Borges2010}
Borges C~V~S, Hor-Meyll M, Huguenin J~A~O and Khoury A~Z 2010 {\em Phys. Rev.
  A\/} {\bf 82}(3) 033833

\bibitem{Ndagano2017}
Ndagano B, Perez-Garcia B, Roux F~S, McLaren M, Rosales-Guzm\'{a}n C, Zhang Y,
  Mouane O, Hernandez-Aranda R~I, Konrad T and Forbes A 2017 {\em Nature
  Phys.\/} {\bf 13} 397--402

\bibitem{Toppel2014}
T{\"o}ppel F, Aiello A, Marquardt C, Giacobino E and Leuchs G 2014 {\em New J.
  Phys.\/} {\bf 16} 073019

\bibitem{Ndagano2018}
Ndagano B, Nape I, Cox M~A, Rosales-Guzm\'{a}n C and Forbes A 2018 {\em J.
  Light. Technol.\/} {\bf 36} 292--301

\bibitem{Dudley2013}
Dudley A, Li Y, Mhlanga T, Escuti M and Forbes A 2013 {\em Opt. Lett.\/} {\bf
  38} 3429--3432

\bibitem{Chen2014}
Chen S, Zhou X, Liu Y, Ling X, Luo H and Wen S 2014 {\em Opt. Lett.\/} {\bf 39}
  5274--5276 \urlprefix\url{http://ol.osa.org/abstract.cfm?URI=ol-39-18-5274}

\bibitem{Wang2007}
Wang X~l, Ding J, Ni W~j, Guo C~s and Wang H~t 2007 {\em Opt. Lett.\/} {\bf 32}
  3549--3551

\bibitem{Maurer2007}
Maurer C, Jesacher A, F{\"{u}}rhapter S, Bernet S and Ritsch-Marte M 2007 {\em
  New J. Phys.\/} {\bf 9} 78

\bibitem{Moreno2012}
Moreno I, Davis J~A, Hernandez T~M, Cottrell D~M and Sand D 2012 {\em Opt.
  Express\/} {\bf 20} 364--376

\bibitem{SPIEbook}
Rosales-Guzm\'{a}n C and Forbes A 2017 {\em How to shape light with spatial
  light modulators\/} SPIE.SPOTLIGHT (SPIE Press)

\bibitem{Rosales2017}
Rosales-Guzm\'{a}n C, Bhebhe N and Forbes A 2017 {\em Opt. Express\/} {\bf 25}
  25697--25706

\bibitem{Ren2015}
Ren Y~X, Lu R~D and Gong L 2015 {\em Annalen der Physik\/} {\bf 527} 447--470

\bibitem{Mitchell2016}
Mitchell K~J, Turtaev S, Padgett M~J, \v{C}i\v{z}m\'{a}r T and Phillips D~B
  2016 {\em Opt. Express\/} {\bf 24} 29269--29282

\bibitem{Scholes2019}
Scholes S, Kara R, Pinnell J, Rodr{\'\i}guez-Fajardo V and Forbes A 2019 {\em
  Optical Engineering\/} {\bf 59} 1 -- 12

\bibitem{Gong2014}
Gong L, Ren Y, Liu W, Wang M, Zhong M, Wang Z and Li Y 2014 {\em J. Appl.
  Phys.\/} {\bf 116} 183105

\bibitem{Yao-Li2020}
Yao-Li, Hu X~B, Perez-Garcia B, Bo-Zhao, Gao W, Zhu Z~H and Rosales-Guzm{\'a}n
  C 2020 {\em Applied Physics Letters\/} {\bf 116} 221105

\bibitem{Zhaobo2020}
Zhao B, Hu X~B, Rodr\'iguez-Fajardo V, Forbes A, Gao W, Zhu Z~H and
  Rosales-Guzm\'an C 2020 {\em Applied Physics Letters\/} {\bf 116} 091101
  (\textit{Preprint} \eprint{https://doi.org/10.1063/1.5142163})
  \urlprefix\url{https://doi.org/10.1063/1.5142163}

\bibitem{Rosales2020}
Rosales-Guzm{\'a}n C, Hu X~B, Selyem A, Moreno-Acosta P, Franke-Arnold S,
  Ramos-Garcia R and Forbes A 2020 {\em Scientific Reports\/} {\bf 10} 10434

\bibitem{Selyem2019}
Selyem A, Rosales-Guzm\'an C, Croke S, Forbes A and Franke-Arnold S 2019 {\em
  Phys. Rev. A\/} {\bf 100}(6) 063842

\bibitem{Manthalkar2020}
Manthalkar A, Nape I, Bordbar N~T, Rosales-Guzm\'{a}n C, Bhattacharya S, Forbes
  A and Dudley A 2020 {\em Opt. Lett.\/} {\bf 45} 2319--2322

\bibitem{Zhaobo2019}
Zhao B, Hu X~B, Rodr\'{i}guez-Fajardo V, Zhu Z~H, Gao W, Forbes A and
  Rosales-Guzm\'{a}n C 2019 {\em Opt. Express\/} {\bf 27} 31087--31093

\bibitem{Beijersbergen1993}
Beijersbergen M, Allen L, van~der Veen H and Woerdman J 1993 {\em Optics
  Communications\/} {\bf 96} 123--132

\bibitem{Oneil2000}
O'Neil A~T and Courtial J 2000 {\em Optics Communications\/} {\bf 181} 35--45

\bibitem{Davis1994}
Davis J~A and Cottrell D~M 1994 {\em Opt. Lett.\/} {\bf 19} 496--498

\bibitem{RosalesGuzman2013Airy}
Rosales-Guzm{\'{a}}n C, Mazilu M, Baumgartl J, Rodr{\'{\i}}guez-Fajardo V,
  Ramos-Garc{\'{\i}}a R and Dholakia K 2013 {\em Journal of Optics\/} {\bf 15}
  044001

\bibitem{MartinezFuentes2018}
Fuentes J~L~M and Moreno I 2018 {\em Opt. Express\/} {\bf 26} 5875--5893

\bibitem{Mirhosseini2013}
Mirhosseini M, Magana-Loaiza O~S, Chen C, Rodenburg B, Malik M and Boyd R 2013
  {\em Opt. Express\/} {\bf 21}

\bibitem{Hu2020}
Hu X~B, Dong M~X, Zhu Z~H, Gao W and Rosales-Guzm{\'a}n C 2020 {\em Scientific
  Reports\/} {\bf 10} 199

\bibitem{Goldstein2011}
Goldstein D~H 2011 {\em Polarized light\/} (CRC Press)

\bibitem{Otte2018a}
Otte E and Denz C 2018 {\em Opt. Lett.\/} {\bf 43} 5821--5824

\bibitem{Bandres2004}
Bandres M~A and Guti\'{e}rrez-Vega J~C 2004 {\em Opt. Lett.\/} {\bf 29}
  144--146

\bibitem{Bandres2004b}
Bandres M~A and Guti\'{e}rrez-Vega J~C 2004 {\em J. Opt. Soc. Am. A\/} {\bf 21}
  873--880 \urlprefix\url{http://josaa.osa.org/abstract.cfm?URI=josaa-21-5-873}

\bibitem{Milione2011}
Milione G, Sztul H~I, Nolan D~A and Alfano R~R 2011 {\em Phys. Rev. Lett.\/}
  {\bf 107}(5) 053601

\bibitem{Meier2007}
Meier M, Romano V and Feurer T 2007 {\em Appl. Phys. A\/} {\bf 86} 329--334

\end{thebibliography}
\providecommand{\newblock}{}

\end{document}